\documentclass[a4paper,11pt]{article}
\usepackage{latexsym}
\usepackage{epsfig}
\usepackage[figuresfirst,nomarkers]{endfloat}
\usepackage{caption2}

\newcommand{\beq}{\begin{eqnarray}}
\newcommand{\eeq}{\end{eqnarray}}
\newcommand{\be}{\begin{eqnarray*}}
\newcommand{\ee}{\end{eqnarray*}}

\newcommand{\D}{{\cal D}}
\newcommand{\Pom}{{\hspace{ -0.1em}I\hspace{-0.25em}P}}
\newcommand{\vphot}{{\gamma^*}}

\newcommand{\QQ}{\scriptscriptstyle{Q \bar{Q}}}
\newcommand{\cc}{{c\bar{c}}}
\def\lsim{\raise0.3ex\hbox{$<$\kern-0.75em\raise-1.1ex\hbox{$\sim$}}}
\def\gsim{\raise0.3ex\hbox{$>$\kern-0.75em\raise-1.1ex\hbox{$\sim$}}}

\long\def\symbolfootnote[#1]#2{\begingroup%
  \def\thefootnote{\fnsymbol{footnote}}\footnote[#1]{#2}\endgroup}

\title{{\bf Charmonium dissociation and recombination \\ at RHIC and LHC}}

\author{A.~Capella$^1$, L.~Bravina$^{2,3}$, E.G.~Ferreiro$^4$,\\
  A.B.~Kaidalov$^5$, K.~Tywoniuk\footnote{{\it E-mail address:}
    konrad@fys.uio.no (K.~Tywoniuk).} $^2$, E.~Zabrodin$^{2,3}$}

\date{}

\begin{document}

\vspace{0.5cm}
\maketitle
 
\begin{center}
\small{
  $^1$ Laboratoire de Physique Th\'eorique\symbolfootnote[2]{Unit\'e Mixte de
    Recherche UMR n$^{\circ}$ 8627 - CNRS}, Universit\'e de Paris XI,
  B\^atiment 210, \\
  91405 Orsay Cedex, France
  \par
  $^2$ Department of Physics, University of Oslo\\
  0316 Oslo, Norway \\
  \par
  $^3$ Institute of Nuclear Physics, Moscow State University\\
  RU-119899 Moscow, Russia\\
  \par
  $^4$ Departamento de F{\'\i}sica de Part{\'\i}culas, Universidad de
  Santiago de Compostela, \\
  15782 Santiago de Compostela, Spain
  \par
  $^5$ Institute of Theoretical and Experimental Physics\\
  RU-117259 Moscow, Russia}
\end{center}

\begin{abstract}
Charmonium production at heavy-ion colliders is considered within the
comovers interaction model. The formalism is extended by including
possible secondary $J/\psi$ production through recombination and an
estimate of recombination effects is made with no free parameters involved. The comovers interaction model also
includes a comprehensive treatment of initial-state nuclear effects,
which are discussed in the context of such high energies. With these
tools, the model properly describes the centrality and the rapidity dependence of
experimental data at RHIC energy, $\sqrt{s} = 200$ GeV, for both {\it Au+Au} and {\it Cu+Cu} collisions. Predictions for LHC, $\sqrt{s} = 5.5$ TeV,
are presented and the assumptions and extrapolations involved are discussed.
\end{abstract}
\vskip 3 truecm

\noindent LPT Orsay 07-125 \par
\noindent December 2007

\newpage

\section{Introduction}
Disentangling effects related to the production of charmonium in
hadronic collisions has been a major task for both experimentalists and
theoreticians for the last two decades. The discovery of $J/\psi$
suppression, with respect to heavy lepton pair
production, in proton-nucleus ({\it pA}) collisions has been interpreted as a result
of the multiple scattering of a $\cc$ pair escaping the nuclear
environment -- the so-called nuclear absorption. 
Moreover, in high
energy nucleus-nucleus ({\it A+A})
collisions one hopes to achieve such large temperatures and
densities that a new state of deconfined QCD matter, the quark-gluon plasma
(QGP), is produced. It was suggested that, in the
presence of a QGP, the charmonium yield would be further
suppressed due to color Debye screening \cite{Matsui86}.
Indeed, such an anomalous, compared to absorption, suppression
was observed in {\it Pb+Pb} collisions at top SPS energy
\cite{Alessandro05}. Alternatively, the SPS experimental results can also
be described in terms of final state interactions of the $\cc$
pairs with the dense medium created in the collision, the so-called
comovers interaction model (CIM). This model does not assume thermal
equilibrium and, thus, does not use thermodynamical concepts. Within
this model the SPS experimental data can be 
reproduced with an effective dissociation cross section $\sigma_{co} =
0.65$ mb \cite{Armesto99}.

The theoretical extrapolations to collider energies are guided
mainly by two trends. On the one hand, models assuming a deconfined phase
during the collision pointed to the growing importance of secondary
$J/\psi$ production due to recombination of $\cc$ pairs in the plasma.
The total amount of $\cc$ pairs is assumed to be created
in hard interactions during the early stages of the collision. Then,
either using kinetic theory and solving rate equations for the
subsequent dissociation and recombination of charmonium
\cite{Thews01,Grandchamp02}, or assuming statistical coalescence at
freeze-out \cite{Braun00,Andronic03,Kostyuk03}, one obtains the final
$J/\psi$ yield. These models predict a disappearance of the charmonium
suppression with rising collision energy as the lifetime of the plasma
phase is expected to grow accordingly. See, however, \cite{Yan06}
where thermal equilibrium is not assumed and charmonium dissociation
is quite large at RHIC. On the other
hand, the CIM with only
dissociation of $J/\psi$'s predicts \cite{Capella05} a stronger suppression at RHIC
than at SPS due to a larger density of produced soft particles in the
collision. It also predicts a stronger suppression at $y=0$ (where the comovers density is maximal) than at forward rapidities.

In this context, measurements of $J/\psi$ production in {\it Au+Au}
collisions at RHIC, $\sqrt{s} = 200$ GeV, gave 
interesting although surprising results - the suppression at
mid-rapidity was on the same level as at SPS \cite{PHENIX07,Leitch07}.
This was also the case for {\it Cu+Cu} collisions at the
same energy \cite{Cianciolo05}.
Furthermore, the suppression at forward rapidity in {\it Au+Au} collisions
was stronger than at mid-rapidity. The latter feature was not
seen for the much smaller collision system created in
{\it Cu+Cu} collisions.

The CIM is based on the well known gain and loss differential
equations in transport theory. The introduction of a recombination
term is actually required for detailed balance. So far it had not been
introduced in the model just because its effect was assumed to be
small. The aforementioned RHIC results prompt us to a careful
evaluation of the effect of the recombination (gain) term. If its
effect is not negligible it will increase the final $J/\psi$ yield.
Moreover, this increase will be larger at $y=0$ than at forward
rapidities due to the narrow rapidity distribution of charm, and thus
complies with RHIC data.

In the present work we will extend the CIM by allowing
recombination of $\cc$ pairs into secondary $J/\psi$'s. We will
estimate this effect using the density of charm in proton-proton ({\it
  pp}) collisions at the 
same energy and at various rapidities. Therefore, the model does not
involve any additional parameters. 
In Section~\ref{sec:model} we present the details of the model; both
nuclear effects related to the initial state at high energy and final state
interactions are described. We calculate recombination effects
at mid- and forward rapidities in {\it Au+Au} and {\it Cu+Cu} collisions
at RHIC energy $\sqrt{s} = 200$ GeV in Section~\ref{sec:rhic}.
The suppression, found without any tuning of the parameters in the
model, is in good agreement with the data. 
In Section~\ref{sec:lhc} we make predictions for {\it Pb+Pb}
collisions at LHC, $\sqrt{s} = 5.5$ TeV, and discuss uncertainties
related to this extrapolation. Finally, conclusions and final remarks
are given in Section~\ref{sec:conclusions}.

\section{Description of the model}\label{sec:model}
In this section, we give a short and updated description of the
CIM which has been used to make predictions for $J/\psi$ production in
{\it A+A} collisions at RHIC \cite{Capella05}. 
This model contains a comprehensive treatment of initial-state nuclear
effects, such as nuclear shadowing and nuclear absorption, and final state
interactions with the co-moving matter. Here we will extend the
model as follows
\begin{itemize}
\item update nuclear shadowing for hard production of charmonium
  calculated from parameterization of diffractive gluon
  distribution function \cite{Aktas06}
\item extend nuclear absorption and its energy dependence to the whole
  kinematically allowed region
\item include the possible recombination of $\cc$ into secondary
  $J/\psi$'s, i.e. the gain term in the differential rate
  equation of the model.
\end{itemize}
The latter effect is the main novel feature that will enable us to
calculate the suppression pattern at RHIC, and make further
predictions for the upcoming heavy-ion runs at LHC.
Nuclear effects in nucleus-nucleus collisions are usually expressed through
the so-called nuclear modification factor, $R^{J/\psi}_{AB}
(b)$, defined as the ratio of the $J/\psi$ yield in {\it A+A} and
{\it pp} scaled by the
number of binary nucleon-nucleon collisions, $n(b)$. We have then
\beq \label{eq:ratioJpsi}
R^{J/\psi}_{AB}(b) \;&=&\;
\frac{\mbox{d}N^{J/\psi}_{AB}/\mbox{d}y}{n(b)
  \,\mbox{d}N^{J/\psi}_{pp}/\mbox{d}y} \nonumber \\ 
\;&=&\; \frac{\int\mbox{d}^2s \, 
  \sigma_{AB}(b) \, n(b,s) \, S_{J/\psi}^{sh}(b,s) \, S^{abs}(b,s) \, S^{co}(b,s)
}{\int \mbox{d}^2 s \, \sigma_{AB} (b) \, n(b,s)} \;,
\eeq
where $\sigma_{AB}(b) = 1 - \exp [-\sigma_{pp}\, AB\, T_{AB}(b)]$, 
$T_{AB}(b) = \int\mbox{d}^2s T_A(s)T_B(b-s)$ is the nuclear overlapping
function and $T_A(b)$ is obtained from Woods-Saxon nuclear densities
\cite{Jager74}. In eq.~(\ref{eq:ratioJpsi}),
\beq
\label{eq:nbin}
n(b,s) \;=\; \sigma_{pp} AB \, T_A(s)\, T_B(b-s)/\sigma_{AB}(b)\;,
\eeq
where upon integration over $\mbox{d}^2s$ we obtain the number of
binary nucleon-nucleon collisions at impact parameter $b$, $n(b)$.

The three additional factors in the numerator of
eq.~(\ref{eq:ratioJpsi}), $S^{sh}$, $S^{abs}$ and $S^{co}$, denote the
effects of shadowing, nuclear absorption (both initial-state effects) and interaction with
the co-moving matter (final-state effect), respectively. They will be
defined below.

\subsection{Initial-state nuclear effects}
The effects related to particle production in hadronic collisions off
nuclei are often called initial-state effects and have been
extensively treated in the literature in the case of $J/\psi$
production at different energies \cite{Boreskov91,Kharzeev97,Vogt05}.
Since the origin and relevance of these effects is still under debate,
we will present a short and updated discussion of them at both low
and ultra-relativistic energies.

At low energies the primordial spectrum of particles created in
scattering off a nucleus is mainly altered by {\bf (i)} interactions 
with the nuclear matter they traverse on the way out to the detector and
{\bf (ii)} energy-momentum conservation.
The first effect is called
nuclear absorption and is usually parameterized within a probabilistic
Glauber model. 
The latter is inferred from the rapidity dependence of the spectra.
For {\it A+A} collisions, these effects can be combined into the following
suppression factor
\beq
\label{eq:Sabs}
S^{abs} = \frac{\left[1 - \exp(-\xi(x_+) \sigma_{\QQ} AT_A(b)) \right]\,
  \left[1 - \exp (-\xi(x_-) \sigma_{\QQ} B T_B(b-s))
  \right]}{\xi(x_+)\xi(x_-) \sigma_{\QQ}^2 AB \, T_A(s) T_B(b-s)} \;,
\eeq
where $\xi(x_\pm) = (1-\epsilon) + \epsilon x_\pm^\gamma$ determines both
absorption and energy-momentum conservation.\footnote{We
recall that $x_\pm = (\sqrt{x_F^2 -4 M^2 /s} \pm x_F)/2$.} In
\cite{Boreskov91} it has been found that $\gamma=2$, $\epsilon=0.75$ and
$\sigma_{\QQ} = 20$ mb give a good description of data. This
corresponds to $\sigma_{abs} = 5$ mb at mid-rapidity which is in
agreement with other studies. 

Several novel features are expected to appear at high energies due to
the change in the space-time picture of particle production.
First of all, the form of eq.~(\ref{eq:Sabs}) will change due to
coherence effects \cite{Braun98}.  Motivated by results from deuteron-gold collisions at
$\sqrt{s} = 200$ GeV \cite{Adler06},  it has recently been realized that
this change should be accompanied by a transformation of $\xi
(x_\pm)\sigma_{\QQ}$ into $\epsilon \, x^{\gamma}_\pm \sigma_{\QQ} $.
In this way $\sigma_{abs} = 0$ at high energies
\cite{Capella06,Arsene07}. Nevertheless, $S^{abs} \not= 1$ for $x_\pm$
not too small, due to energy-momentum conservation. This leads to a
model for the high-energy transition in which the factors in the
numerator of 
eq.~(\ref{eq:Sabs}) are changed according to
\beq
\label{eq:SabsHE}
\frac{1 \,-\, e^{-\xi (x_\pm)\,\sigma_{\QQ} A T_A(b)}}{\xi
  (x_\pm) \, \sigma_{\QQ}} \;\longrightarrow \; A T_A (b) \, \exp \left[-
\epsilon \sigma_{\QQ} x_\pm^\gamma \, A T_A(b) /2 \right] \;.
\eeq
In the central rapidity region at RHIC, $x_+ \sim
0.025-0.05$ and absorption effects can be discarded.
At $|y| > 0$, $S^{abs} \neq 1$ due to energy-momentum conservation.
This effect is usually neglected in the discussion of {\it A+A}
collisions.

Secondly, coherence effects will lead to nuclear shadowing for both soft
and hard processes at RHIC, and therefore also for the production of
heavy flavor. Shadowing can be calculated
within the Glauber-Gribov theory \cite{Gribov69}, and we will utilize
the generalized Schwimmer model of multiple scattering
\cite{Schwimmer75}. In this case the second suppression factor in
eq.~(\ref{eq:ratioJpsi}) is given by 
\beq
\label{eq:schwimmer}
S^{sh}(b,s,y) \;=\; \frac{1}{1 \,+\, A F(y_A) T_A(s)} \, \frac{1}{1
  \,+\, B F(y_B) T_B (b-s)} \;,
\eeq
where the function $F(y)$ encodes the dynamics of shadowing and will
be discussed in detail below. For a general discussion of nuclear shadowing,
see \cite{Capella99,Armesto06}.

Parameterizations of diffractive structure function from $\vphot N$
scattering as measured at HERA have been utilized in \cite{Armesto03} to
find shadowing for sea quarks. Most recently
gluon shadowing has been calculated in \cite{Tywoniuk07} using recent data
on diffractive gluon density function \cite{Aktas06}. Then shadowing
is governed by
\beq
\label{eq:h1shad}
F(x,Q^2) \;=\; 4\pi \, \int_x^{x^{max}_\Pom} \!\!\!\mbox{d} x_\Pom \,
B(x_\Pom) \, \frac{F_{2\D}^{(3)} \left(x_\Pom, Q^2,\beta \right) }{F_2
  \left( x,Q^2 \right)} \, F_A^2 (t_{min} ) \;, 
\eeq
where $F(x,Q^2)$ is related to $F(y)$ in eq.~(\ref{eq:schwimmer})
through kinematical relations. For gluon fusion, $x = m_T \exp(\pm
y)/\sqrt{s}$. We put $x^{max}_\Pom = 0.1$, where
shadowing is expected to disappear.
In eq.~(\ref{eq:h1shad}), $F_2(x,Q^2)$ is the structure function for a nucleon,
$F_{2\D}^{(3)}(x_\Pom,Q^2,\beta)$ is the t-integrated diffractive structure
function of the nucleon, $B(x_\Pom)$ is the slope parameter of the
distribution, and $F_A (t_{min})$ is the nuclear form factor where $t_{min}
\approx - m_N^2 x_\Pom^2$. Equation~(\ref{eq:schwimmer}) determines shadowing
for quarks (antiquarks) in
nuclei \cite{Armesto03}.
For gluons the same expressions have been used with the substitutions:
$F_{2\D}^{(3)} \left(x_\Pom,Q^2,\beta \right) \rightarrow F^g_\Pom
\left(x_\Pom,Q^2,\beta \right)$, $F_2 \left(x,Q^2 \right) \rightarrow
x g\left(x,Q^2 \right)$, $F^g_\Pom$ and $g$ represent gluon distributions in the Pomeron,
measured in diffractive deep inelastic scattering, and in the
proton, respectively. We use this information to calculate shadowing
for $J/\psi$ assuming that it originates from a
color octet $\cc$ pair produced in a hard gluon fusion process with
$Q = m_T = 4$ GeV. The same value of $Q$ is used for open charm production.

Since experiments at HERA mostly deal with hard diffraction and their
parameterizations are quite uncertain for $Q^2 < 4$ GeV$^2$, the
density of comovers, mostly low-$p_\perp$ particles, will be calculated in the spirit
of the model presented in \cite{Capella99,Capella01}, where
shadowing corrections are given without free parameters
in terms of the triple-Pomeron coupling determined from diffraction data. Then, in
eq.~(\ref{eq:schwimmer})
\beq
\label{eq:simpleshad}
F(y_A) \;=\; C_{sh} \left[\exp\left(\Delta Y_{max}^A \right) \,-\, \exp
  \left(\Delta Y_{min}^A \right) \right] \;,
\eeq
where $Y_{min}^A = \ln (R_A m_N/\sqrt{3})$, $\Delta = 0.13$ and $C_{sh} =
0.31$ fm$^2$. The value of $Y_{max}$ depends on the rapidity of the
produced particle $h$, $Y_{max}^A = \ln(s/m_T^2)/2\pm y$ with the
$+(-)$ sign if $h$ is produced in the hemisphere of nucleus $B(A)$. 
$m_T$ is the transverse mass of the produced particle. We use $m_T =
0.4$~GeV at RHIC and 0.5~GeV at LHC. This 
model has been used to correctly predict the centrality dependence of soft
particle production at RHIC \cite{Capella01}.

The energy dependence of these different effects can be
summarized as follows.
Particle production at SPS is dominated by low-energy effects,
i.e. nuclear absorption given by eq.~(\ref{eq:Sabs}) and small
nuclear shadowing, while RHIC already belongs to the high-energy
regime. Nuclear shadowing is non-negligible at mid-rapidity, and the
combined effect of shadowing and energy-momentum conservation, in the
spirit of eq.~(\ref{eq:SabsHE}), should be accounted for at forward
rapidities. At LHC, shadowing will be very strong even at $y=0$
while energy-momentum conservation is a small effect and can be
neglected in most of the kinematics. We will now proceed with the
discussion of final state effects.

\subsection{Dissociation by comovers interaction and recombination}
The CIM was developed to explain both the
suppression of charmonium yields
\cite{Armesto99,Capella05,Brodsky88,Koch90,Capella97,Armesto98,Capella00}
and the strangeness enhancement \cite{Capella95,Capella96} in
nucleus-nucleus collisions at the SPS. Neglecting possible
recombination effects, the rate equation governing the
density of charmonium in the final state, $N_{J/\psi}$, can be written
in a simple form assuming a pure longitudinal expansion of the system and boost invariance.
For an  {\it A+A} collision the
density of $J/\psi$ at a given 
transverse coordinate, $s$, impact parameter $b$, and rapidity is given by
\beq
\label{eq:comovrateeq}
\tau \frac{\mbox{d} N_{J/\psi}}{\mbox{d} \tau} \, \left( b,s,y \right)
\;=\; -\sigma_{co} N^{co}(b,s,y) N_{J/\psi}(b,s,y) \;,
\eeq
where $\sigma_{co}$ is the cross section of charmonium dissociation
due to interactions with the co-moving medium, with density $N^{co}$. It is found
from fits to low-energy experimental data to be $\sigma_{co} = 0.65$
mb \cite{Armesto99}. To incorporate the effects of recombination, we have to include an
additional gain term proportional to the (squared) density of open charm
produced in the collision. Then eq.~(\ref{eq:comovrateeq}) is
generalized to
\beq
\label{eq:recorateeq}
\tau \frac{\mbox{d} N_{J/\psi}}{\mbox{d} \tau} \, \left( b,s,y \right)
\;=\; -\sigma_{co} \left[ N^{co}(b,s,y) N_{J/\psi}(b,s,y) \,-\,
  N_c(b,s,y) N_{\bar{c}} (b,s,y) \right] \;,
\eeq
where we have assumed that the effective recombination cross section
is equal to the dissociation cross section. \footnote{These two
  cross-sections have to be similar but not necessarily equal. We have
  taken the simplest possibility.}. This extension of the model 
therefore does not involve additional parameters.

Equation~(\ref{eq:recorateeq}) cannot be solved analytically. We
approximate its solution as
\beq
\label{eq:fullsupp}
S^{co}(b,s,y) \;=\; \exp \left\{-\sigma_{co}
  \,\left[N^{co}(b,s,y)\,-\, \frac{N_c(b,s,y)
  N_{\bar{c}} (b,s,y)}{N_{J/\psi}(b,s,y)} \right] \, \ln
\left[\frac{N^{co}(b,s,y)}{N_{pp} (0)}\right] \right\} \;,
\eeq
resembling the exact solution of eq.~(\ref{eq:comovrateeq}), since the
first term in the exponent of eq.~(\ref{eq:fullsupp}) is exactly the
survival probability of a $J/\psi$ interacting with comovers
\cite{Capella05}. The density of open and hidden 
charm in {\it A+A} collisions, $N_c,N_{\bar{c}}$ and $N_{J/\psi}$,
respectively, can be computed from their densities in {\it
  pp} collisions as $N_c^{AA} (b,s) = n(b,s)
S_{HQ}^{sh}(b,s)N_c^{pp}$, with similar expression for
$N_{\overline{c}}^{AA}$ and $N_{J/\psi}^{AA}$. Here $n(b,s)$ is 
given by eq.~(\ref{eq:nbin}) and $S_{HQ}^{sh}$ is the shadowing factor
for heavy quark production, given by eq.~(\ref{eq:h1shad}). Then
eq.~(\ref{eq:fullsupp}) becomes
\beq
S^{co}(b,s,y) \;=\; \exp \left\{-\sigma_{co} \,\left[N^{co}(b,s,y)- C(y)
    n(b,s)S_{HQ}^{sh}(b,s) \right] \, \ln
  \left[\frac{N^{co}(b,s,y)}{N_{pp} (0)}\right] \right\}
\eeq
where
\beq
\label{eq:Cratio}
C (y) \;=\; \frac{\left(\mbox{d}N^{\cc}_{pp}/\mbox{d}y
  \right)^2}{\mbox{d} 
  N^{J/\psi}_{pp}/\mbox{d} y} \;=\; \frac{\left(\mbox{d}
    \sigma^{\cc}_{pp}/\mbox{d}y 
  \right)^2}{\sigma_{pp} \,
  \mbox{d}\sigma^{J/\psi}_{pp}/\mbox{d} y} \;.
\eeq
The quantities in the rightmost term in eq.~(\ref{eq:Cratio}) are all
related to {\it pp} collisions at the corresponding energy and can be
taken from experiment or a model (for extrapolation of the
experimental results).
The $\cc$ pairs are mostly in  charmed mesons,
such as $D$ and $D^*$.
With $\sigma_{co}$ fixed from experiments at low energy, where
recombination effects are negligible, the model, formulated above,
should be self-consistent at high energies\footnote{Note, however,
  that $\sigma_{co}$ could change when the energy increases. We do not
  expect this effect to be important and, since we are unable to
  evaluate the magnitude of this eventual change, we shall use the same
  value $\sigma_{co} = 0.65$~mb at all energies.}. 
We expect the effect of recombination to be stronger at
mid- than at forward rapidities. At $y \neq 0$ the recombination term is smaller
(relative to the 
first one) since the rapidity distribution of $D$, $D^*$ is
narrower than the one of comovers.
This will produce a decrease of $R_{AA}^{J/\psi}$ with increasing $y$
which may over-compensate the increase due to a smaller density of
comovers at $y \neq 0$.

The density of comovers is calculated using the dual parton
model \cite{Capella94} together with the proper shadowing correction. Then
\beq
\label{eq:DensComov}
N^{co} (b,s,y) \;=\; N^{co}_{NS} (b,s,y) \, S^{sh}_{ch} (b,s,y) \;,
\eeq
where $S_{ch}^{sh}$ denotes the shadowing for light particles and is
given by eqs.~(\ref{eq:schwimmer}) and (\ref{eq:simpleshad}). The
non-shadowed (NS) multiplicity of comovers is then
\beq
N^{co}_{NS} (b,s,y) \;=\; \frac{3}{2} \, \frac{ \mbox{d} N^{ch}_{NS}
  (b,s,y)}{\mbox{d} y} \;=\; \frac{3}{2} \, \left\{ C_1(b,y) n_A(b,s)
  \,+\, C_2(b,y) n(b,s) \right\} \;,
\eeq
where
\beq
n_A(b,s) \;=\; \frac{AT_A(s)}{\sigma_{AB}(b)} \, \left[ 1 \,-\,
  \exp\left( -\sigma_{pp} BT_B(b-s) \right) \right] \;,
\eeq
and $n(b,s)$ is given by eq.~(\ref{eq:nbin}). The numerical values of the
coefficients $C_1$ and $C_2$ can be found in \cite{Capella05}.
$C_1$ drops with energy while $C_2$ increases, and at LHC $C_1 \approx 0$
while $C_2 \approx 6$ at mid-rapidity. 
Also in eq.~(\ref{eq:fullsupp}) $N_{pp} (0) = \frac{3}{2} (\mbox{d} 
N^{ch}/\mbox{d} y  )^{pp}_{y=0} / \pi R_p^2$, which we
estimate\footnote{$pp$ values at LHC of $dN/dy$ and $\sigma$ are based
  on DPMJET-III calculations. We thank J.~Ranft for providing these
  results.}  to be 
2.24~fm$^{-2}$ at $\sqrt{s} = 200$ GeV and 4.34~fm$^{-2}$ at $\sqrt{s}
= 5.5$ TeV. Finally, $\sigma_{pp}$, taken as its non-diffractive
value, is 34 mb and 59 mb at RHIC and
LHC, respectively.

\section{Charmonium production at RHIC}\label{sec:rhic}
\begin{figure}[t!]
  \begin{center}
    \includegraphics[width=.5\linewidth]{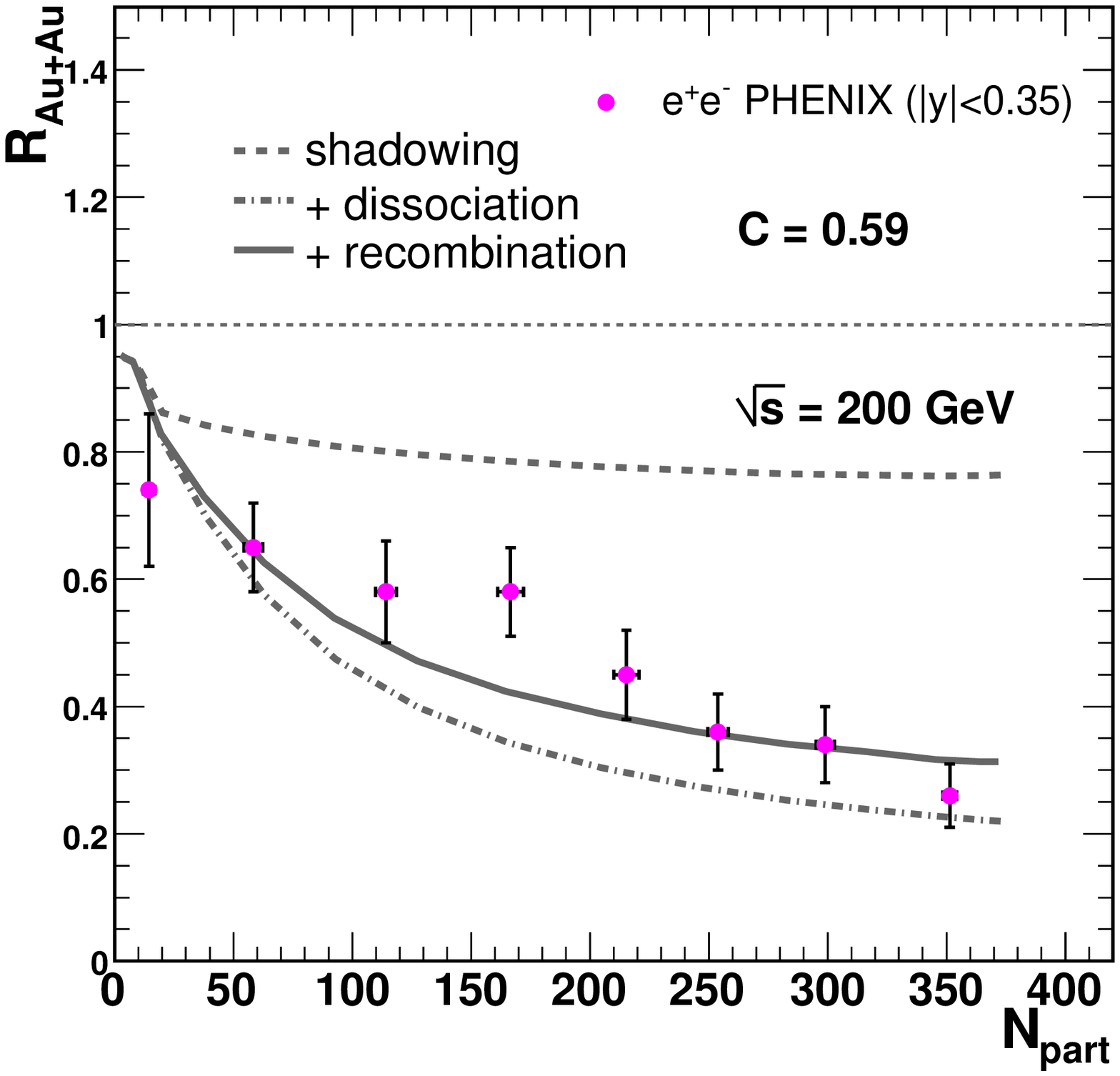}%
    \includegraphics[width=.5\linewidth]{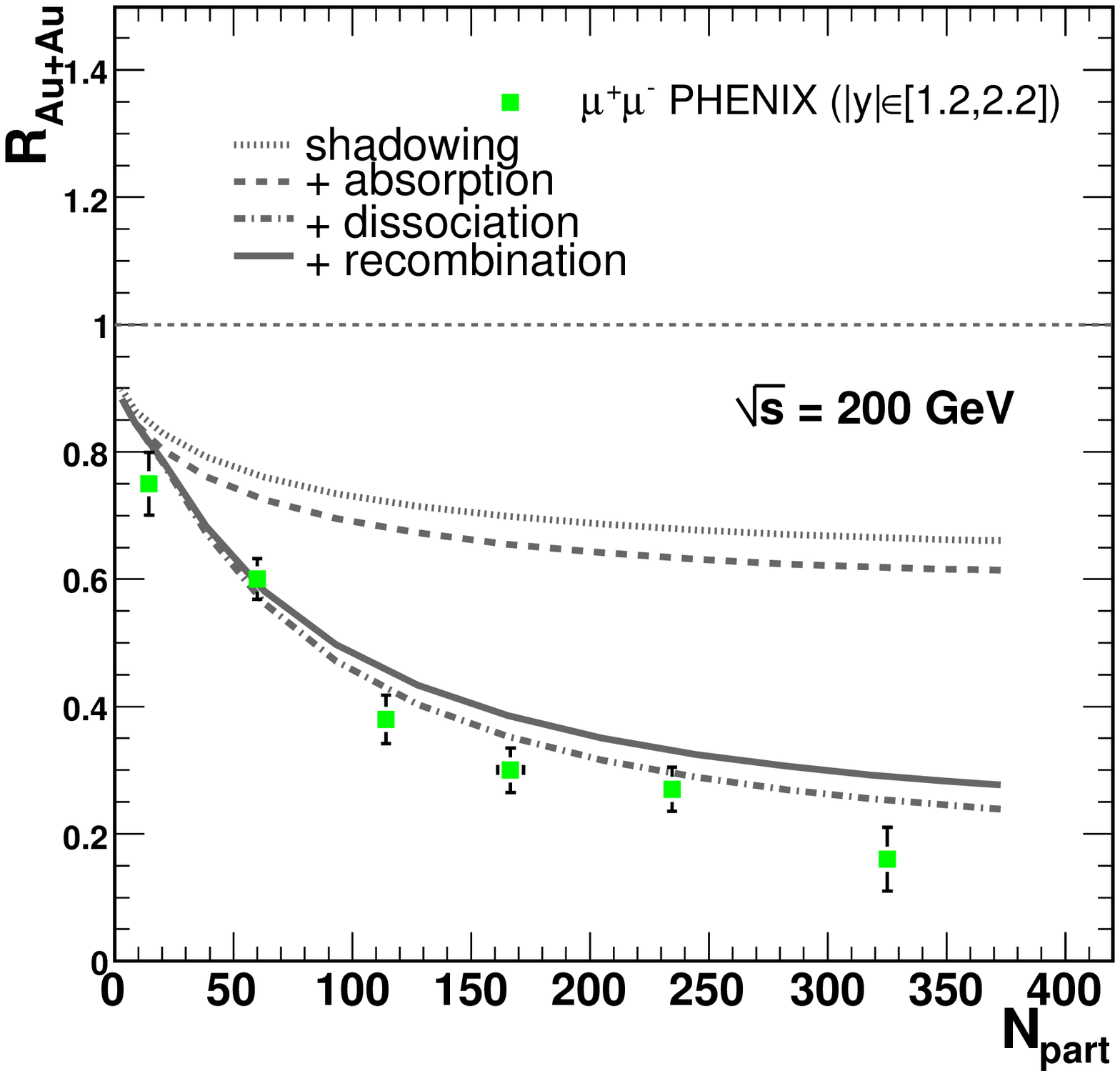}
  \end{center}
  \caption{Results for $J/\psi$ suppression in {\it Au+Au} at RHIC
    ($\sqrt{s} = 200$ GeV) at mid- (left 
    figure), and at forward rapidities
    (right figure). Data are from \cite{PHENIX07}. The solid curves
    are the final results. The dashed-dotted ones are the results
    without recombination ($C = 0$). The dashed line is the total
    initial-state effect. The dotted line in the right figure is the
    result of shadowing. In the left figure the last two lines
    coincide (see main text).} 
  \label{fig:JPSIresultsRHICAuAu}
\end{figure}
At RHIC, the dissociation term alone gives a too strong suppression
compared to experimental data \cite{Capella05}. We therefore proceed
to estimate the effect of recombination. The density of open charm at
mid-rapidity in {\it pp} 
collisions at $\sqrt{s} = 200$ GeV has been reported in
\cite{Adare06} and the most recent measurement of the $J/\psi$ density
in \cite{Adare:2006kf}. 
We present results of experimental measurements in Table~\ref{tbl:pp}.
The semi-leptonic branching ratio for the $J/\psi$ is $B_{ll} =
0.059$. Then, from eq.~(\ref{eq:Cratio}) it follows that $C =
0.59$ at mid-rapidities. One has to keep in mind, however, that the
measured yield of open
charm at RHIC is almost twice as large as predictions from pQCD
\cite{Cacciari05}.
In the left picture of
Fig.~\ref{fig:JPSIresultsRHICAuAu} we present the results of our
model compared to experimental data at mid-rapidity. The different
contributions to $J/\psi$ suppression are shown (see figure caption).
Note that at mid-rapidities the initial-state effect is just the
shadowing. As discussed above nuclear absorption due to
energy-momentum conservation is present at forward rapidities but is
negligibly small at mid-rapidities. 
\begin{table}
  \caption{Open charm and $J/\psi$ production cross sections in {\it
      pp} collisions at $\sqrt{s} = 200$ GeV. Data are taken from
    \cite{Adare06,Adare:2006kf,Adler06forw}.}
  \begin{center}
    \begin{tabular}{c||c|c|c}
      & $\left( \mbox{d} \sigma^{\cc}_{pp} / \mbox{d} y
      \right)_{EXP}$ & $\left( \mbox{d} \sigma^{\cc}_{pp} / \mbox{d} y
      \right)_{PYTHIA}$ & $\left( B_{ll} \, \mbox{d} \sigma^{J/\psi}_{pp}
        / \mbox{d} y \right)_{EXP}$ \\
      \hline
      mid-rap & 123 $\pm$ 12 $\pm$ 45 $\mu$b & & 44.3 $\pm$ 1.4 $\pm$ 5.1
      nb \\
      \hline
      forward & & 70.9 $\pm$ 14 $\mu$b &
      27.61 $\pm$ 0.37 $\pm$ 0.83 nb \\
    \end{tabular}
  \end{center}
  \label{tbl:pp}
\end{table}

Measurements of open charm at forward rapidities \cite{Adler06forw}
have too large systematic errors at the moment.
Therefore, to estimate the density of charm at forward rapidity we 
use results of PYTHIA \cite{Sjostrand01} with parameters and settings as described in
\cite{Adler06forw}. The resulting rapidity distributions can be
described by slightly broadened Gaussian functions. Note, that with this
PYTHIA value, given in Table~1, the ratio of open charm production at
mid- and forward rapidities is very similar to the one measured for
$J/\psi$ production. With these values of $J/\psi$ and open charm
rapidity distributions  we get $C = 0.32$ at forward rapidities. Using
this value of $C$ and a ratio of 1.2 \cite{43r} of the comover
densities $N^{co}$ between $y = 0$ and $y$ forward we obtain the curve
in the r.h.s. of Fig.~1. We note that no free parameters were tuned to
obtain 
these results. Note that, contrary to the results in \cite{Capella05}
with no recombination, the $J/\psi$ suppression at forward rapidity is
somewhat larger that the one at mid-rapidities, in agreement with
experimental data. This is due both to the recombination term and to
the initial-state effects. The latter are stronger for forward
rapidities. They include the effects of energy-momentum conservation,
which were not considered in \cite{Capella05}. 

\begin{figure}[t!]
  \begin{center}
    \includegraphics[width=.5\linewidth]{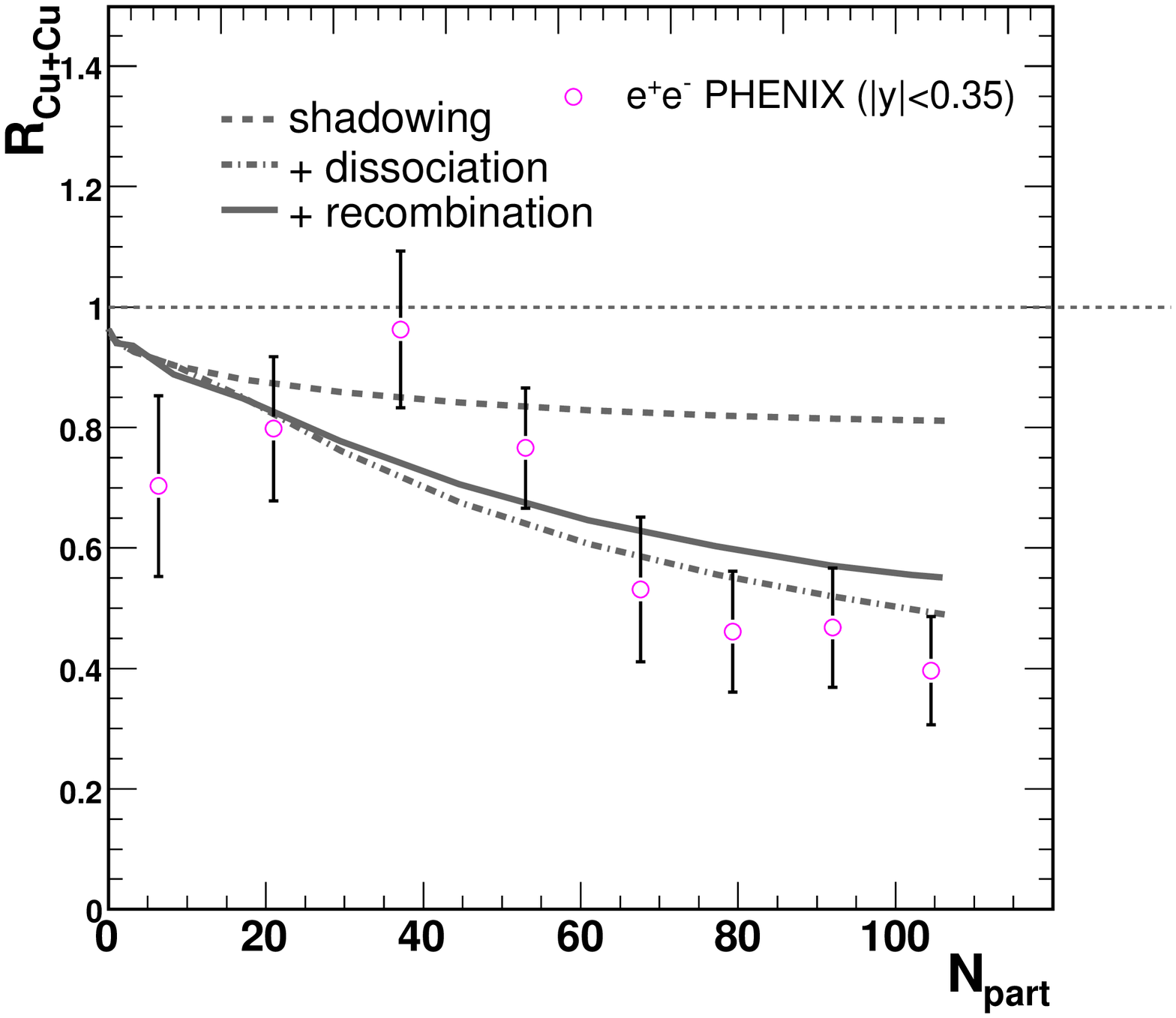}%
    \includegraphics[width=.5\linewidth]{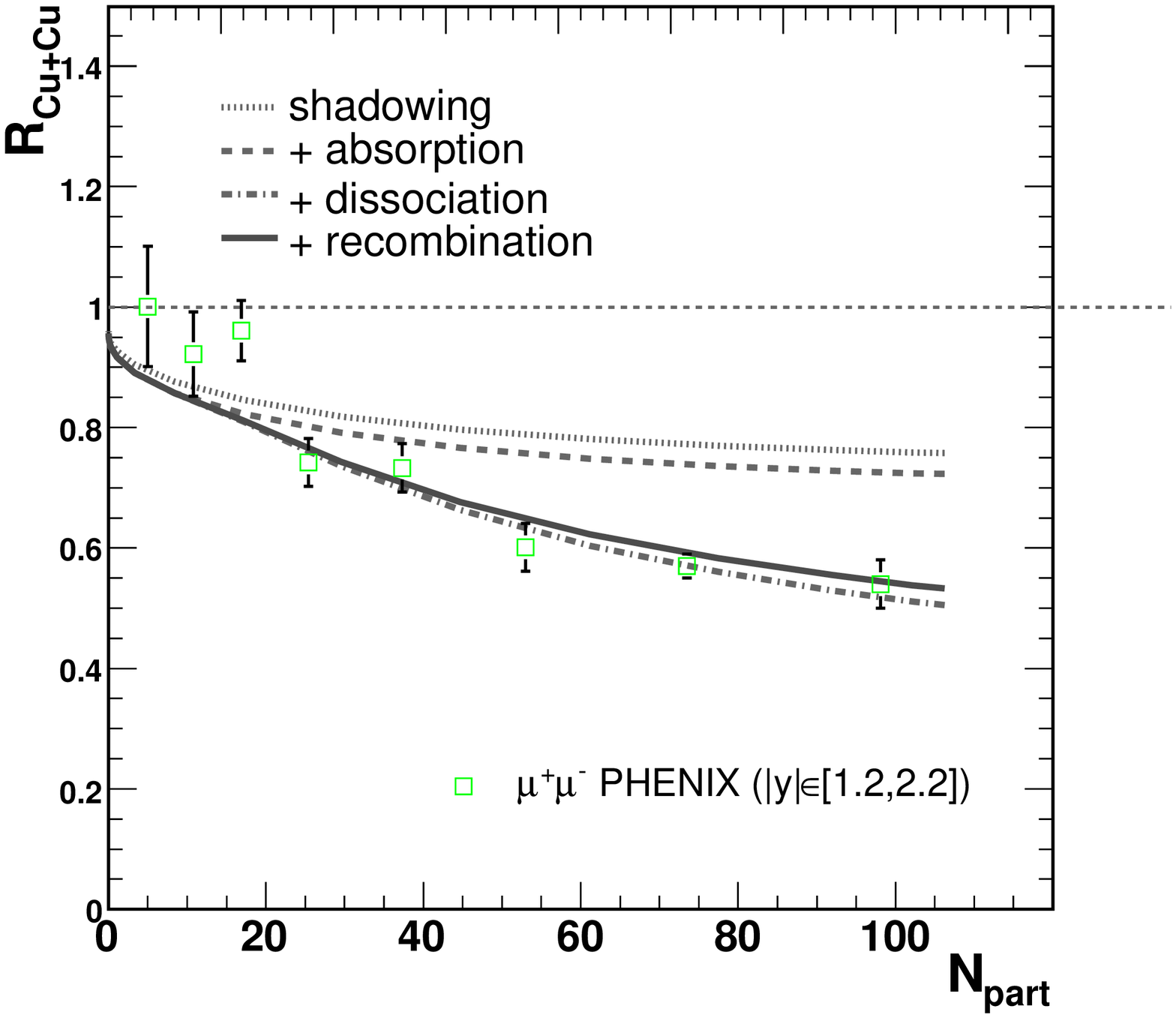}
  \end{center}
  \caption{Results for $J/\psi$ suppression in  {\it Cu+Cu} at RHIC
    ($\sqrt{s} = 200$ GeV), at mid- (left) and 
    at forward rapidities (right). For details, see caption of
    Fig.~\ref{fig:JPSIresultsRHICAuAu}. Data are from
    \cite{Cianciolo05}.}
  \label{fig:JPSIresultsRHICCuCu}
\end{figure}
For consistency, we have also made calculations for the $J/\psi$
suppression in {\it Cu+Cu} 
collisions at RHIC which was reported in \cite{Cianciolo05}, using the
same parameters as above for {\it Au+Au} collisions. The results are
shown in Fig.~\ref{fig:JPSIresultsRHICCuCu}, and are in good agreement
with the experimental data, except for peripheral collisions where the
error bars are quite large.

Concluding, our  procedure gives a reasonable description of data
both at mid- and forward rapidity for different collision systems at
RHIC, in contrast with the recent claim \cite{Linnyk207} that the CIM
fails to reproduce forward rapidity data. We have shown that the
situation is much improved by inclusion of a recombination term.

\section{Predictions for LHC}\label{sec:lhc}
\begin{figure}[t!]
  \begin{center}
    \includegraphics[scale=0.4]{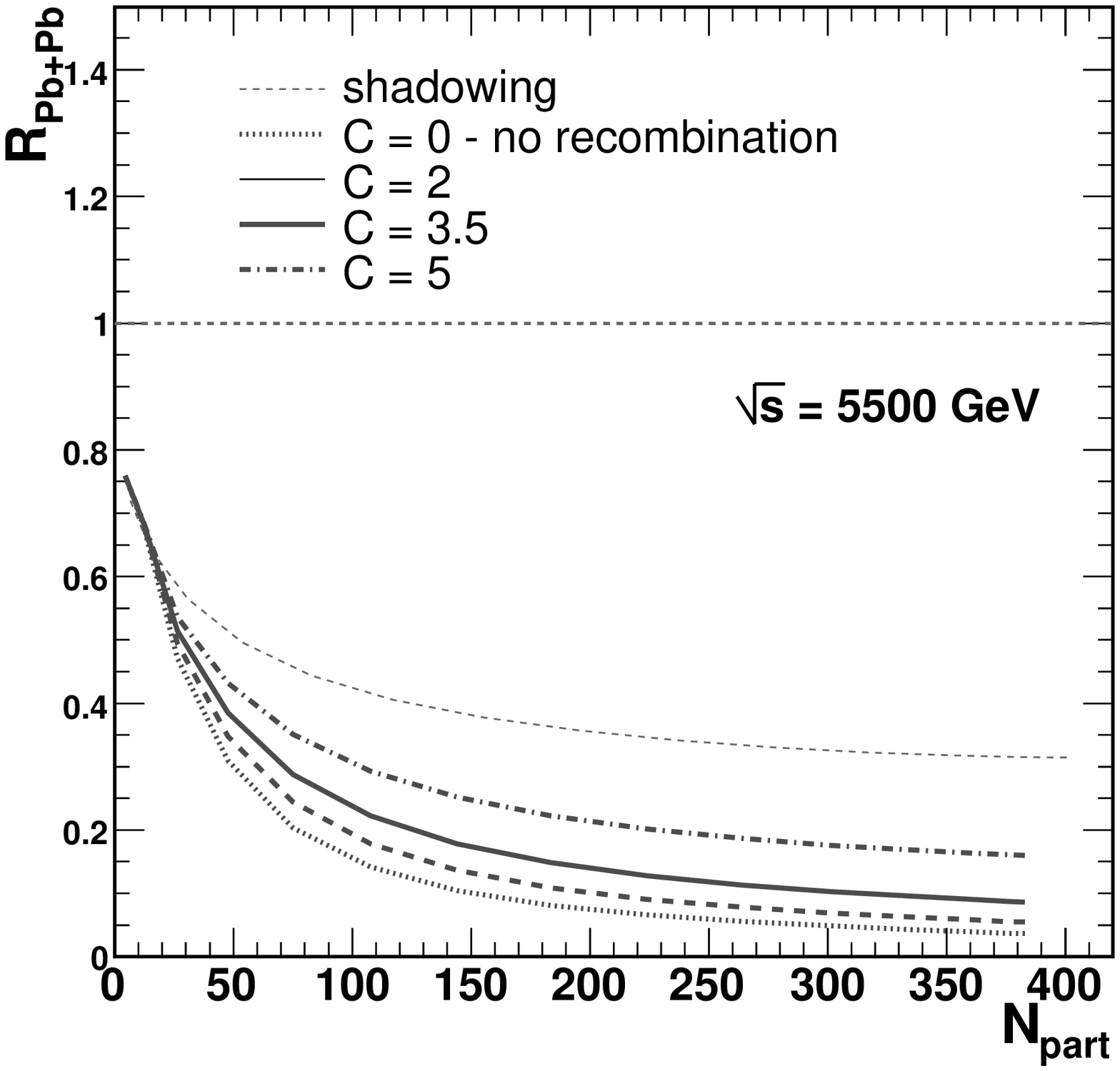}
  \end{center}
  \caption{Results for $J/\psi$ suppression in {\it Pb+Pb} at LHC ($\sqrt{s} = 5.5$
    TeV) at mid-rapidities for different values
    of the parameter $C$. The upper line is the suppression due to initial-state effects (shadowing).} 
  \label{fig:JPSIresultsLHC}
\end{figure}

Based on our previous discussion, it is obvious that recombination
effects will be of crucial importance in {\it Pb+Pb} collisions at LHC
($\sqrt{s} = 5.5$ TeV). Assuming that the energy dependence of open charm and $J/\psi$ in {\it pp} collisions is the same (between RHIC and LHC energies), the energy dependence of the
parameter $C$
will be that  
of $\sigma^{\cc}_{pp}/\sigma_{pp}$. The total and differential cross
section for charm can be calculated using perturbative techniques
\cite{Cacciari05,Vogt07}. The calculations for low energies
are in agreement with data, yet predictions for RHIC and Tevatron energies
are lower than the data. Therefore, the extrapolation to LHC is quite
uncertain. If we parameterize the energy dependence of open charm
production as $\sigma^{\cc} \propto s^\alpha$, with $\alpha 
= 0.3$ and use the values of non-diffractive $\sigma_{pp}$ given at
the end of Section 2 we obtain $C = 2.5$ at LHC -- a value about four
times larger than the corresponding one at RHIC. In view of that we
consider that realistic values of $C$ at LHC are of the range 2 to 3.
In Fig.~\ref{fig:JPSIresultsLHC} we have calculated the $J/\psi$
suppression at LHC 
for several values of C, including the case of absence of
recombination effects ($C=0$).
Although the density of charm grows substantially from RHIC to LHC,
the combined effect of initial-state shadowing and comovers
dissociation appears to overcome the effect of parton recombination.
This is in sharp contrast with the findings of \cite{Braun00}, where
a strong enhancement of the $J/\psi$ yield with increasing centrality
was predicted.

It is clear that a better theoretical control is needed on the
various factors that are included in $C$. As discussed in Section~2
the comovers cross section
$\sigma_{co}$ can vary a little with energy. But the most important
theoretical input at the moment 
is the energy dependence of the total charm cross section in ${\it
  pp}$ collisions.

\section{Conclusion}\label{sec:conclusions}

In this work we have incorporated the effects of 
recombination of $\cc$ pairs into $J/\psi$ in the comovers interaction
model. These effects are 
negligible at low energies (SPS) due to the low density of open
charm. This model does not assume thermal equilibrium of the matter
produced in the collision and
includes a comprehensive treatment of initial-state effects, such as
shadowing, nuclear absorption and energy-momentum conservation.

We estimate the magnitude of the recombination term from
$J/\psi$ and open charm yields in  {\it
  pp} collisions at RHIC.
Without any adjustable parameters, the centrality and
rapidity dependence of experimental data is reproduced both for {\it
  Au+Au} and {\it Cu+Cu} collisions.

Finally, we make predictions for future measurements of $J/\psi$ in
{\it Pb+Pb} collisions at LHC. In our approach, the magnitude of the
recombination effect is 
controlled by the total charm cross section in ${\it pp}$ collisions, and therefore our
predictions are strongly dependent on input from theoretical models at
these energies. For a reasonable choice of parameters, we predict that
the suppression observed at RHIC and lower energies will 
still dominate over the recombination effects.
This is  due to the large density of
comovers and to the strong initial-state suppression at these
ultra-relativistic energies.

\section*{Acknowledgments}
This work was supported by the Norwegian
Research Council (NFR) under contract No.~166727/V30,
RFBF-06-02-17912, RFBF-06-02-72041-MNTI, INTAS 05-103-7515, grant of
leading scientific schools 845.2006.2 and Federal
Agency on Atomic Energy of Russia.


\begin{thebibliography}{99}
\bibitem{Matsui86} T.~Matsui, H.~Satz, Phys. Lett. B {\bf 178}, 416
  (1986)

\bibitem{Alessandro05} B.~Alessandro, et al., Eur.\ Phys.\ J.\  C {\bf
    39}, 335 (2005)

\bibitem{Armesto99} N.~Armesto, A.~Capella, E.G.~Ferreiro, Phys.
  Rev. C {\bf 59}, 395 (1999)

\bibitem{Thews01} R.~Thews, M.~Schroedter, J.~Rafelski, Phys. Rev.
  C {\bf 63}, 054905 (2001)

\bibitem{Grandchamp02} L.~Grandchamp, R.~Rapp, Nucl. Phys. A {\bf
    709}, 415 (2002)

\bibitem{Braun00} P.~Braun-Munzinger, J.~Stachel, Phys. Lett. B {\bf
    490}, 196 (2000)

\bibitem{Andronic03} A.~Andronic, P.~Braun-Munzinger, K.~Redlich,
  J.~Stachel, Phys. Lett. B {\bf 571}, 36 (2003)

\bibitem{Kostyuk03} A.P.~Kostyuk, M.I.~Gorenstein, H.~St\"ocker,
  W.~Greiner, Phys. Rev. C {\bf 68}, 041902 (2003)

\bibitem{Yan06} L.~Yan, P.~Zhuang, N.~Xu, Phys.\ Rev.\ Lett.\
  {\bf 97}, 232301 (2006)
  
\bibitem{Capella05} A.~Capella, E.G.~Ferreiro, Eur. Phys. J. C {\bf
    42}, 419 (2005)

\bibitem{PHENIX07} A.~Adare, et al., Phys.\ Rev.\ Lett.\  {\bf 98},
  232301 (2007) 

\bibitem{Leitch07} M.J.~Leitch, J.\ Phys.\ G {\bf 34}, S453 (2007),
  \texttt{nucl-ex/0701021} 

\bibitem{Cianciolo05} V.~Cianciolo, AIP Conf.\
  Proc.\  {\bf 842}, 41 (2006), \texttt{nucl-ex/0601012}

\bibitem{Aktas06} A.~Aktas, et al., Eur.\ Phys.\ J.\  C {\bf 48}, 715
  (2006); {\it ibid.} {\bf 48}, 749 (2006)

\bibitem{Jager74} C.W.~De Jager, H.~De Vries and C.~De Vries, Atom.\
  Data Nucl.\ Data Tabl.\  {\bf 14}, 479 (1974)

\bibitem{Boreskov91} K.G.~Boreskov, A.B.~Kaidalov, in {\it Proceedings of
  the XXVIth Rencontre de Moriond, Savoie, 1991}, edited by J. Tran
  Thanh Van, (Editions Frontieres, 1991)\\
  K.G.~Boreskov, A.~Capella, A.B.~Kaidalov, J.~Tran~Thanh~Van,
  Phys.\ Rev.\ D {\bf 47}, 919 (1993)

\bibitem{Kharzeev97} D.~Kharzeev, C.~Louren\c{c}o, M.~Nardi, H.~Satz,
  Z. Phys. C {\bf 74}, 307 (1997)

\bibitem{Vogt05} R.~Vogt, Phys. Rev. C {\bf 71}, 054902 (2005)

\bibitem{Braun98} M.A.~Braun, C.~Pajares, C.A.~Salgado, N.~Armesto,
  A.~Capella, Nucl. Phys. B {\bf 509}, 357 (1998)

\bibitem{Adler06} S.S.~Adler, et al., Phys.\ Rev.\ Lett.\  {\bf 96},
  012304 (2006)
  
\bibitem{Capella06} A.~Capella, E.G.~Ferreiro, \texttt{hep-ph/0610313}

\bibitem{Arsene07} I.C.~Arsene, L.~Bravina, A.B.~Kaidalov,
  K.~Tywoniuk, E.~Zabrodin, Phys.\ Lett.\ B (in press),
  \texttt{arXiv:0711.4672 [hep-ph]}\\
  K.~Tywoniuk, et al., J.\ Phys.\ G (in press), \texttt{arXiv:0712.2382 [hep-ph]}
  
\bibitem{Gribov69} V.N.~Gribov, Sov.\ Phys.\ JETP {\bf 29}, 483 (1969);
  {\it ibid.} {\bf 30}, 709 (1970); {\it ibid.} {\bf 26}, 414 (1968)

\bibitem{Schwimmer75} A.~Schwimmer, Nucl.\ Phys.\ B {\bf 94}, 445 (1975)

\bibitem{Capella99} A.~Capella, A.~Kaidalov, J.~Tran Thanh Van,
  Heavy Ion Phys. {\bf 9}, 169 (1999), \texttt{hep-ph/9903244}

\bibitem{Armesto06} N.~Armesto, J.\ Phys.\ G {\bf 32}, R367 (2006)

\bibitem{Armesto03} N.~Armesto, A.~Capella, A.B.~Kaidalov,
  J.~Lopez-Albacete, C.A.~Salgado, Eur. Phys. J. C {\bf 29}, 531 (2003)

\bibitem{Tywoniuk07} K.~Tywoniuk, I.~Arsene, L.~Bravina, A.~Kaidalov,
  E.~Zabrodin, Phys.\ Lett.\ B {\bf 657}, 170 (2007); Eur.\ Phys.\
  J.\ C {\bf 49}, 193 (2007)

\bibitem{Capella01} A.~Capella, D.~Sousa, Phys. Lett. B {\bf 511},
  185 (2001)

\bibitem{Brodsky88} S.~Brodsky, A.H.~Mueller, Phys. Lett. B {\bf
    206}, 685 (1988)
 
\bibitem{Koch90} B.~Koch, U.~Heinz, J.~Pitsut, Phys. Lett. {\bf
    243}, 149 (1990)

\bibitem{Capella97} A.~Capella, A.~Kaidalov, A.~Kouider Akil,
  C.~Gerschel, Phys.\ Lett.\  B {\bf 393}, 431 (1997) 

\bibitem{Armesto98} N.~Armesto, A.~Capella, Phys. Lett. B {\bf
    430}, 23 (1998)

\bibitem{Capella00} A.~Capella, E.G.~Ferreiro, A.B.~Kaidalov,
  Phys. Rev. Lett. {\bf 85}, 2080 (2000)

\bibitem{Capella95} A.~Capella, Phys.\ Lett.\  B {\bf 364}, 175 (1995)

\bibitem{Capella96} A.~Capella, A.~Kaidalov, A.~Kouider Akil,
  C.~Merino, J.~Tran Thanh Van, Z.\ Phys.\  C {\bf 70}, 507 (1996)\\
  A.~Capella, C.A.~Salgado, D.~Sousa, Eur.\ Phys.\ J.\ C {\bf 30},
  111 (2003)

\bibitem{Capella94} A.~Capella, U.~Sukhatme, C.I.~Tan, J.~Tran
  Thanh Van, Phys.\ Rept.\  {\bf 236}, 225 (1994)

\bibitem{Adare06} A.~Adare, et al., Phys.\ Rev.\ Lett.\  {\bf 97},
  252002 (2006)

\bibitem{Adare:2006kf} A.~Adare, et al.,
  Phys.\ Rev.\ Lett.\  {\bf 98}, 232002 (2007)

\bibitem{Cacciari05} M.~Cacciari, P.~Nason, R.~Vogt, Phys. Rev.
  Lett. {\bf 95}, 122001 (2005)

\bibitem{Adler06forw} S.S.~Adler, et al.,
  Phys.\ Rev.\  D {\bf 76}, 092002 (2007)

\bibitem{Sjostrand01}
  T.~Sjostrand, P.~Eden, C.~Friberg, L.~Lonnblad, G.~Miu, S.~Mrenna,
  E.~Norrbin, Comput.\ Phys.\ Commun.\  {\bf 135}, 238 (2001) 
  
\bibitem{43r} 
  R.~Debbe, AIP Conf.\ Proc.\  {\bf 870}, 707 (2006),
  \texttt{nucl-ex/0608048}

\bibitem{Linnyk207} O.~Linnyk, E.L.~Bratkovskaya, W.~Cassing,
  H.~St\"ocker, Phys.\ Rev.\  C {\bf 76}, 041901 (2007)

\bibitem{Vogt07} R.~Vogt, \texttt{arXiv:0709.2531 [hep-ph]}

\end{thebibliography}
\end{document}